\documentclass[final]{elsart}
\usepackage[latin1]{inputenc}
\usepackage{epsfig}
\usepackage{color}
\usepackage{amsmath}
\usepackage{amsfonts}
\usepackage{amssymb}
\usepackage{wasysym}
\usepackage{graphicx}
\journal{Applied Energy}

\begin{document}

\begin{frontmatter}

\title{On cycle-to-cycle heat release variations in a simulated spark ignition heat engine}

\author{P.L. Curto-Risso\thanksref{uruguay}}
   \address{Departamento de F\'isica Aplicada, Universidad de Salamanca, 37008 Salamanca, Spain}
   \thanks[uruguay]{Also at: Instituto de Ingeniería Mecánica y Producción Industrial, Universidad de la República, 11300 Montevideo, Uruguay.}
    \ead{pcurto@fing.edu.uy}

\author{A. Medina\corauthref{cor1}\thanksref{bejar}}
    \address{Departamento de F\'isica Aplicada, Universidad de Salamanca, 37008 Salamanca, Spain}
    \corauth[cor1]{Corresponding author. Tel. +34 923 29 44 36; fax: +34 923 29 45 84.}
    \thanks[bejar]{Also at: ETSII de B\'ejar, Universidad de Salamanca, 37700 B\'ejar, Salamanca, Spain.}
    \ead{amd385@usal.es}

\author{A. Calvo Hern\'andez\thanksref{IUFFYM}}    
\address{Departamento de F\'isica Aplicada, Universidad de Salamanca, 37008 Salamanca, Spain}
 \thanks[IUFFYM]{Also at: IUFFYM, Universidad de Salamanca, 37008, Salamanca, Spain.}
\ead{anca@usal.es}

\author{L. Guzm\'an-Vargas}
\address{Unidad Profesional Interdisciplinaria en Ingenier{\'\i}a y Tecnolog{\'\i}as Avanzadas, Instituto Polit\'ecnico Nacional, Av. IPN No.~2580, L. Ticom\'an, M\'exico D.F. 07340, M\'exico}
\ead{lguzmanv@ipn.mx}

\author{F. Angulo-Brown}
\address{Departamento de F{\'\i}sica, Escuela Superior de F{\'\i}sica y Matem\'aticas, Instituto Polit\'ecnico Nacional, Edif. No. 9 U.P. Zacatenco, 
M\'exico D.F. 07738, M\'exico}
\ead{angulo@esfm.ipn.mx}

\date{\today}

\begin{abstract}
The cycle-by-cycle variations in heat release for a simulated
spark-ignited engine are analyzed within a turbulent combustion model in terms of some basic parameters: the characteristic
length of the unburned eddies entrained within the flame front, a characteristic turbulent speed, and the
location of the ignition kernel. The evolution of the simulated time
series with the fuel-air equivalence ratio, $\phi$, from lean mixtures to over stoichiometric
conditions, is examined and compared with previous experiments. Fluctuations on the
characteristic length of unburned eddies are found to be essential to simulate heat release cycle-to-cycle variations and recover experimental results. Relative to the non-linear analysis of the system,
it is remarkable that at  fuel ratios around $\phi\simeq 0.65$, embedding and surrogate procedures show that the dimensionality of the system is small.
\end{abstract}

\end{frontmatter}
\maketitle

\section{Introduction}\label{s1}

Nowadays,  primary objectives for vehicle manufacturers are the reduction of fuel consumption as well as the
reduction of emissions. Among the several solutions to be adopted are fast combustion, lean burn, variable valve
timing, gasoline direct injection and some others~\cite{fontana2009}. Combustion in spark-ignition engines in lean burn conditions
can lead to cycle-to-cycle variations (CV) that are responsible for higher emissions and limit the practical levels of lean-fueling.
Better models for simulating the dynamic character of CV could lead to new design and control alternatives for lowering emissions
and enlarging the actual operation limits.

The cyclic variation experienced by spark ignition engines is essentially a
 combustion problem which is affected by many engine and
operating variables like fuel properties, mixture composition near
spark plug, charge homogeneity, ignition,
exhaust dilution, etc~\cite{heywood.intro}.  Physical sources of CV arise from the merging of at least the
following elements~\cite{ozdor94,abdi2007}:

\begin{enumerate}
\item Gases mixture motion and turbulences during combustion.

\item The amounts of fuel, air, and residual or recirculated exhaust gas in the cylinder.

\item Homogeneity of the mixture composition, specially in the vicinity of the spark plug.

\item Details of the spark discharge, like breakdown energy and the initial flame kernel random position.
\end{enumerate}

Not all these factors are equally important, and some others were not included in the list. It is usually argued that the first item in the list is the leading effect. The flow motion in any piston engine is unsteady, having an average pattern and a fluctuating velocity component. Nevertheless, it is not easy to select an adequate turbulence parameter. 
During the engine operation not all the burnt gases are expelled from the combustion chamber in the
exhaust stroke. Some fraction is mixed with the next incoming air and fuel mixture in the following intake stroke.
So, when combustion approaches the lean flammability limit, very small modifications due to recycled gases in the mixture composition
or temperature can greatly affect the next combustion event causing  a highly non-linear cycle-to-cycle coupling process.

Besides experimental and thermodynamic theoretical studies~\cite{rocha06}, computer
simulations~\cite{pedro2008,Curto-Risso2009} are useful tools which provide
complementary understanding of the complex physical mechanisms
involved in the operation of a real engine. Several previous studies~\cite{shen1996,abdi2007} tried to reproduce experimental CV by incorporating explicit models for ignition, combustion chemistry, flame evolution, and turbulence. In these models some input parameters are varied in each cycle in such a way that cycle-to-cycle combustion also changes, and so the pressure and temperature in the cylinder and subsequently heat release, power output and efficiency.  Abdi Aghdam \emph{et al.}~\cite{abdi2007} reproduced satisfactory fits to pressure and flame radius data by performing quasi-dimensional computer simulations and acting on the rms turbulent velocity.

In regard to  CV phenomenology we mention the following experimental and theoretical works, many of them related to the non-linear dynamics involved in these variations. Letelier \textit{et al.}~\cite{letelier97} reported experimental
results for the variation of the in-cylinder pressure versus crank
angle for a four-cylinder spark heat engine. By reconstructing  the phase space and building the
Poincar\'e sections it was concluded that CV is
not governed by a chaotic process but rather by a superimposition of
a non-linear deterministic dynamics with a stochastic component.
In the works by Daw \textit{et al.}~\cite{daw96,daw98}, a physically
oriented simple mathematical model was reported for the spark
ignited engine. The fuel/air mass ratio in a cycle
was modeled as stochastic through random noise and
non-linear deterministic coupling between consecutive engine cycles.
 The
predicted CV trends for heat release versus fuel-air ratio were
compared with experimental results by analyzing bifurcation plots
and return maps for different parameters of the model  (the mean
 fuel-air ratio, its standard deviation,
and the fraction of the un-reacted air and fuel remaining in the
cylinder). Also, an analysis of the temporal
irreversibility in time-series measurements was presented by using a
symbolic approach to characterize the noisy dynamics~\cite{daw00}.

Litak and coworkers~\cite{litak07,sen08,sen2010b} have reported extensive work relative to cycle-to-cycle
variations of peak pressure, peak pressure location and heat release
in a four-cylinder spark ignition engine in terms on spark advance
angle and different torque loadings under near stoichiometric
conditions. The observed qualitative change in combustion was
analyzed using different statistical methods as return maps, recurrence plots~\cite{litak07,litak08}, correlation coarse-grained entropy
~\cite{litak05} and, recently, multi-fractal techniques~\cite{sen08,sen2010},
and the so-called multi-scale entropy (MSE) (or sample
entropy)~\cite{litak09}: an improved statistical tool to account
for complexity measure in time series with multiple temporal or
spatial scales.

Along this line, a main goal
of this paper is to analyze the effect of some combustion
parameters and their consequences on cycle-to-cycle variations of heat release for fuel-air equivalence ratios
ranging from very lean mixtures to stoichiometric conditions. To get this we developed a
quasi-dimensional computer simulation that incorporates turbulent
combustion and  a detailed analysis of the involved chemical reaction that includes in a natural way the presence of recycled exhaust gases
in the air-fuel fresh mixture and valves overlapping. The model, previously
developed and validated~\cite{pedro2008}, allows for a systematic study
of the influence on CV of three basic combustion parameters: the
characteristic length of the unburned eddies, the characteristic turbulent speed,
and the location of the ignition site. Our simulations correctly reproduce  the basic statistical parameters for the fuel ratio dependence
of experimental heat release time series~\cite{daw98,sen2010}.
Moreover, we shall employ some usual techniques from non-linear data analysis in order to characterize our results and compare them with those from other authors: first-return maps, bifurcation-like plots, and embedding and surrogate techniques to obtain the correlation dimension.

The paper is organized in the following way. Section~II is devoted
to briefly describe the basic elements of our quasi-dimensional
computer simulation with special emphasis on combustion. In Sec.~III
we analyze the simulation results for heat release time series at different values of the fuel/air equivalence ratio and compare with previous experimental results. Sec.~IV is devoted to analyze the dimensionality of the system by means of embedding and surrogate procedures. Finally, in Sec. V we discuss our results
and summarize the main conclusions of the paper.

\section{Simulated model: basic framework}\label{s2}

We make use of a \emph{quasi-dimensional} numerical simulation of a
monocylindrical spark ignition engine, previously described and
validated~\cite{pedro2008}. Opposite to \emph{zero-dimensional} (or thermodynamical)
models where all thermodynamic variables are averaged over a finite volume and an empirical correlation is used to approximate the combustion process, within the quasi-dimensional scheme explicit differential equations are considered to solve combustion under the hypothesis of a spherical flame front. In our simulation model two ordinary differential equations are solved in each time step (or each crankshaft angle) for the pressure and the temperature of the gases inside the cylinder. This set of coupled equations (that are explicitly written in,~\cite{pedro2008} , ~\cite{Curto-Risso2009}, and ~\cite{pedro_tesis})  are valid for any of the different steps of the engine evolution (intake, compression, power stroke, and exhaust) except combustion. For this stage a two-zone model  discerning between unburned ($u$) and burned ($b$) gases
that are separated by an adiabatic flame front with negligible volume is considered. Thus, the temperature equation splits in equations for $T_u$ and $T_b$. Moreover, differential equations giving the evolution of the unburned and burned mass of gases should be solved. Next  we detail the main assumptions of the combustion model considered.

\subsection{Combustion process}

Combustion is the most important process in cycle-by-cycle
variations. For simulating combustion we assume the turbulent
quasi-dimensional model (sometimes called \emph{eddy-burning} or \emph{entrainment} model) proposed by Blizard and
Keck~\cite{keck_simp,blizard_1974} and improved by Beretta \emph{et
al.}~\cite{beretta}, by considering a diffusion term in the burned mass equation as we shall see next.
The model starts from the idea that during flame propagation not all the mass inside the approximately
spherical flame front is burned, but there exist unburned eddies of
typical length $l_t$. Thus, a set of coupled differential equations
for the time evolution of total mass within the flame front,
$m_\textrm{e}$ (unburned eddies plus burned gas) and burned mass,
$m_\textrm{b}$, is solved:
\begin{equation}
 \dot{m}_{b}=\rho_{u} A_{f} S_{l}+\frac{m_{e}-m_{b}}{\tau_{b}}
\label{mb}
\end{equation}
\begin{equation}
 \dot{m}_{e}=\rho_{u} A_{f}\left[u_{t} \left(1-e^{t/\tau_{b}} \right) +S_{l}\right]
\label{me}
\end{equation}
where $u_t$ is a characteristic velocity at which unburned gases pass through the flame front, $\rho_{u}$ is the unburned gas density, and $A_{f}$ is the flame front area. $\tau_b$ is a characteristic time for the combustion of the entrained eddies, $\tau_b=l_t/S_l$ and $S_l$ is the laminar combustion speed that is determined from its reference value, $S_{l,0}$, as~\cite{heywood.laminar}:
\begin{equation}
 S_{l}=S_{l,0}\left( \frac{T_{u}}{T_{\textrm{ref}}}\right) ^{\alpha}\left( \frac{p}{p_{\textrm{ref}}}\right) ^{\beta}\left(1-2.06\,y_ {r}^{0.77} \right)
\label{llama.laminar}
\end{equation}
where $y_{r}$ is the mole fraction of residual gases in the chamber
and $\alpha$ and $\beta$ are functions of the \emph{fuel/air
equivalence ratio}, $\phi$~\cite{bayraktar_LPG}. The reference laminar burning speed,
$S_{l,0}$, at reference conditions ($T_{\textrm{ref}}$, $p_{\textrm{ref}}$) is
obtained from Gülder's model~\cite{gulder,bayraktar_LPG}. Because of the reasons that we shall detail later, it is important to stress that the laminar speed,
apart from the thermodynamic conditions, depends on the fuel/air
equivalence ratio and on the mole fraction of gases in the chamber
after combustion. Thus, Eq.~(\ref{llama.laminar}) shows the coupling between combustion dynamics
and the residual gases in the cylinder after the previous combustion event. In other words laminar burning
speed depends on the memory of the chemistry of combustion.
From Eq.~(\ref{me}) it is easy to see that $\tau_b$ determines the
time scale associated either to the laminar diffusive evolution of the
combustion at a velocity $S_l$ or to the rapid convective turbulent
component associated to $u_t$.

In order to numerically solve the differential equations for
masses (Eqs.~(\ref{mb}) and~(\ref{me})), it is necessary to determine $A_{f}$, $u_{t}$, and $l_{t}$.
The flame front area, $A_f$, is calculated from the flame front
radius, considering a spherical flame propagation in a disc shaped
combustion chamber~\cite{blizard_1974}. The radius is calculated through the enflamed
volume, $V_f$, following the procedure described by
Bayraktar~\cite{bay_mat_model},
\begin{equation}
V_{f}=\frac{m_{b}}{\rho_{b}}+\frac{m_{e}-m_{b}}{\rho_{u}}
\end{equation}
The flame front area depends on the relative position of the kernel of
combustion respect to the cylinder center, $R_c$, which have its nominal value at the spark plug
position, but convection at early times can produce significant displacements.
For $u_{t}$ and $l_{t}$, Beretta \emph{et al.}~\cite{beretta} have derived empirical correlations
in terms of the ratio between the fresh mixture density at ambient
conditions, $\rho_i$, and the density of the unburned gases mixture
inside the combustion chamber, $\rho_u$:
\begin{equation}
u_{t}=0.08\,\bar{u}_{i}\left(\frac{\rho_{u}}{\rho_{i}}\right)^{\frac{1}{2}}
\label{ut}
\end{equation}
\begin{equation}
 l_{t}=0.8\, L_{v,\textrm{max}} \left(\frac{\rho_{i}}{\rho_{u}}\right)^{\frac{3}{4}}
\label{lt}
\end{equation} where $\bar{u}_{i}$ is the mean inlet gas speed, and $L_{v,\textrm{max}}$ the maximum valve lift.
Equations~(\ref{mb}) and (\ref{me}) must be solved
simultaneously together with the set of differential equations for
temperature and pressure~\cite{pedro2008}. Respect to the end of
combustion, we assume that if exhaust valve opens before the
completion of combustion, it finishes at the moment of the opening.

Among the parameters that determine the development of combustion,
there are three essential ones: the characteristic length of eddies
$l_{t}$ (associated to the characteristic time, $\tau_b$), the turbulent entrainment velocity $u_{t}$ (essential for the slope of $m_e(t)$ during the fastest stage of combustion), and also $A_f$
or the flame radius center $R_{c}$ that gives the size and geometry of the
flame front (and thus influences all process). Therefore, all
these parameters could have strong influence on cycle-by-cycle
variations.

In order
to calculate the heat release during combustion we apply the first
principle of thermodynamics for open systems separating heat
release, $\delta Q_{r}$, internal energy
variations associated to temperature changes, $d U$,
work output, $\delta W$, and heat transfers from the
working fluid (considered as a mixture of ideal gases) to the
cylinder walls, $\delta Q_\ell$:
\begin{equation}
 \delta Q_{r}=d U+\delta W+\delta Q_\ell
\end{equation}
where internal energy and heat losses include terms associated to
either unburned or burned gases: $U=m_{u} c_{v,u} T_{u} + m_{b}
c_{v,b} T_{b}$ and $\delta Q_\ell= \delta
Q_{\ell,u}+\delta Q_{\ell,b}$. All these terms can be derived in terms of time or of the crankshaft angle. Net heat
release during the whole combustion period is calculated from the
integration of heat release variation during that period. A particular model to evaluate heat transfer through cylinder walls should be chosen before the numerical obtention of $Q_r$.
Units throughout the paper are international system of units except when explicitly mentioned.

\subsection{Heat transfer}

Among the different empirical models that can be found in the
literature for heat transfer between the gas and cylinder internal
wall we shall utilize that developed by
Woschni~\cite{Woschni1967},
\begin{equation}
\frac{\dot{Q_\ell}}{A_{t}\left(T-T_{w} \right)
}=129.8\,p^{0.8}w^{0.8}B^{-0.2}T^{-0.55}
\end{equation} where $A_{t}$ is the instantaneous heat transfer area, $p$ is the pressure inside the cylinder, $T$ the gas temperature, $T_{w}$ the cylinder internal wall temperature, $B$ the cylinder bore, and $w$ the corrected mean piston speed, that is calculated as follows,
\begin{equation}
w=C_{1}v_{P}+C_{2}\frac{V_{dt}T_{r}}{p_{r}V_{r}}\left(p-p_{m}
\right)
\end{equation}
$C_{1}$ and $C_{2}$ are constant parameters~\cite{stone_heat}, $v_{P}$ is the
mean piston speed, $V_{dt}$ is the  maximum displaced volume, $p_{m}$ is a motoring pressure~\cite{stone_heat}, and the subscript
$r$ refers to conditions immediately after inlet valve is closed. All units are in the
international  system of units, except pressures which are in bar.

\subsection{Chemical reaction}

In order to solve the chemistry and the energetics of combustion we
assume the unburned gas mixture as formed by a standard fuel for spark ignition engines (iso-octane,
$\text{C}_8\text{H}_{18}$), air, and exhaust gases. Exhaust chemical
composition appears directly as calculated from solving combustion.
The considered chemical reaction is,
\begin{align}
& \left( 1-y_{r}\right)\left[  \text{C}_8\text{H}_{18}+\alpha
(\text{O}_{2}+3.773\text{N}_{2})\right]+  \nonumber\\ &
 y_{r}\biggl[\beta_{r} \text{CO}_{2} +\gamma_{r} \text{H}_{2}\text{O} + \mu_{r} \text{N}_{2} + \nu_{r} \text{O}_{2} +
\varepsilon_{r} \text{CO} + \delta_{r} \text{H}_{2}+\nonumber\\ &
 +\xi_{r} \text{H} +\kappa_{r} \text{O} +\sigma_{r} \text{OH} +\varsigma_{r} \text{NO}\biggr]\nonumber  \longrightarrow  \\
&\beta \text{CO}_{2} + \gamma \text{H}_{2}\text{O} + \mu \text{N}_{2} + \nu \text{O}_{2} + \varepsilon
\text{CO} + \delta \text{H}_{2}+\nonumber\\ &
+ \xi \text{H} + \kappa \text{O} + \sigma \text{OH} + \varsigma \text{NO}
\end{align} 
where the subscript $r$ refers to residual. We make use of the subroutine developed by Ferguson~\cite{Ferguson.EQ} (but including residual gases among the reactants) to solve combustion and calculate exhaust composition.
Our model does not consider traces of $\text{C}_{8}\text{H}_{18}$ in combustion
products, but in the energy release we actually take into account
combustible elements as $\text{CO}$, $\text{H}$, or $\text{H}_{2}$. The thermodynamic
properties of all the involved chemical species are obtained from
the constant pressure specific heats, that are taken as
$7$-parameter temperature polynomials~\cite{CEA}.

As a summary of the theoretical section it is important to note that in our dynamical system  the coupled ordinary differential equations for pressure and temperature are in turn coupled with two other ordinary differential equations for the evolution of the masses during combustion. So, globally this leads to a system of differential equations where apart from several parameters (mainly arising  from the geometry of the cylinder, the chemical reaction and the models considered for other processes), there is a large number of variables: pressure inside the cylinder, $p$, temperatures of the unburned and burned gases, $T_u$, and $T_b$, and  masses of the unburned and burned gases, $m_u$ and $m_b$. All these variables evolve with time or with the crankshaft angle. So, up to this point, our dynamical model  is a quite intricate deterministic system with those time dependent variables.

\section{Results}\label{s3}

In this section we present the numerical results obtained from the simulations outlined above and taking the numerical values for the cylinder geometry from Beretta \emph{et al.}~\cite{beretta} and for a fixed engine speed of $109$ rad/s. Details on some running parameters of the computations can be found in~\cite{pedro2008}.

The computed results for heat release, $Q_{r}$, for each of the first $200$
cycles are presented in Fig.~1(a) for a particular value of the fuel-air equivalence ratio, $\phi=1.0$. This time evolution was obtained directly from the solution of the deterministic set of differential equations. The curve clearly shows, after a transitory period, a regular evolution that does not match with previous experimental studies of cycle-to-cycle variability on heat release~\cite{daw98,litak07}. For other fuel ratios the curves obtained present different levels of variability but never display the typical experimental fluctuations.
In order to analyze these time series we show in Table I for several fuel ratio values, some usual statistical
parameters: the average value $\mu$, the standard deviation $\sigma$, the
coefficient of variance, $COV=\sigma/\mu$, the skewness
$S=\sum_{i=1}^N(x_i-\mu)^3/[(N-1)\sigma^3]$, and the kurtosis
$K=\sum_{i=1}^N(x_i-\mu)^4/[(N-1)\sigma^4]$. $S$ is a measure
of the lack of symmetry in such a way that negative (positive)
values imply the existence of left (right) asymmetric tails longer
than the right (left) tail. $K$ is a measure of whether the data are
peaked ($K> 3$) or flat ($K< 3$) relative to a normal distribution.

With the objective to recover the experimental results on cyclic variability we have checked the influence of incorporating a stochastic component in any of the physically relevant parameters of the combustion model.
We first analyze the influence of the characteristic
length $l_t$ and velocity $u_t$ keeping the
location of ignition at the spark plug position
(\emph{i.e.}, $R_c = 25\times 10^{-3}$ m~\cite{beretta}). Although
both parameters can be considered as independent in our model (see
below), we assume that they are linked by the empirical
relations~(\ref{ut}) and~(\ref{lt}). We fitted the experimental results by Beretta~\cite{beretta} for $l_t$ considered as a random variable to a log-normal probability distribution, $LogN(\mu_{\log l_t},\sigma_{\log l_t})$, around the nominal value $l_t^0=0.8
L_{v,max}(\rho_i/\rho_u)^{3/4}$ [Eq.~(\ref{lt})] with standard deviation
$\sigma_{\log l_t}=0.222$ and mean $\mu_{\log l_t}=\log
(l_t^0)-(\sigma_{\log l_t}^2/2)$. Then, in each cycle $u_t$ is obtained from Eq.~(\ref{ut}) through the empirical densities ratio.

Table II for several values of $\phi$ ranging from very lean mixtures to over stoichiometric, show the characteristics of the time evolution of the heat release when the stochastic component on $l_t$ has been introduced. Figure~1(b) displays a time series for the stochastic computations at least qualitatively much similar to the experimental ones than the deterministic for $\phi=1.0$. It is important to note that it is difficult to perform a direct quantitative comparison with experiments, because of the large number of geometric and working parameters of the engine~\cite{pedro2008}, usually not specified to the full extent in the experimental publications.

From a comparison between Tables I and II is clear that average values are of course similar, but the standard deviation and covariance are much smaller in the deterministic case. The behavior of skewness is subtle, it only could be concluded that globally in the deterministic case values for $S$ are closer to zero (that would be the value corresponding to a Gaussian distribution). For the stochastic simulation kurtosis in the fuel ratio interval between $\phi=0.6$ and $0.8$ is much higher than in the deterministic case, that is a sign of more peaked distributions. Figure~2 shows the evolution of heat release with the number of cycles for fuel ratios between $0.5$ and $1.1$. A careful inspection shows that at low and intermediate fuel-ratios distributions are quite asymmetric with tails displaced to the left. In other words there exist remarkable poor combustion or misfire events. This is quantified by the evolution with $\phi$ of the skewness compiled in Table II: it takes high negative values in the interval $\phi=0.5-0.8$. These results are in accordance with the experimental ones by Sen \emph{et al.}~\cite{sen2010} (see Fig.~1 in that paper).

We represent in Fig.~3 the evolution with the fuel-air equivalence ratio of the statistical parameters contained in Tables I and II in order to perform a direct comparison with previous experimental results. Although the experiments by Daw \emph{et al.}~\cite{daw98,sen2010} were performed for a real V8 gasoline engine with a different cylinder geometry that the considered in our simulations, the dependence on $\phi$ of the standard deviation, covariance, skewness and kurtosis is very similar, although, of course, vertical scales in simulations and experiments are different. It makes no sense, to compare the mean values of heat release because the different shape and size of the real and the simulated engine. It is clear from the figure that at intermediate fuel ratios, around, $\phi=0.7$, skewness also reproduces a pronounced minimum and kurtosis a sharp maximum. Nevertheless, our deterministic heat release computations do not reproduce neither the evolution with $\phi$ shown by experiments nor the magnitude of those statistical parameters.

Figure~4 represents the first-return maps, $Q_{r,i+1}$ versus $Q_{r,i}$, for the same values of $\phi$ obtained from the deterministic simulation (dark points) and also from the stochastic computations (color points).  Return maps  when $l_t$ does not have a stochastic component, for any value of fuel ratio, seem noisy unstructured kernels. When $l_t$ is considered as stochastic a rich variety of shapes for return maps is found, depending on the fuel ratio.
From the stochastic return maps in this figure and the statistical results in Table II we stress the
following points. First,
 at high values of $\phi$ ($0.9-1.1$) variations of heat release behave as small
\textit{noisy} spots characteristics of small amplitude
distributions with asymmetric right tails and slightly more peaked near the
mean than a Gaussian distribution. For $\phi=1.0$ and $1.1$ those noisy spots are partially overlapped. These results are very similar to the experimental ones by Daw \emph{et al.} (see Fig.~1 in~\cite{Green1999}) for different engines and those by Litak \emph{et al.} (see Fig.~4 in~\cite{litak07}). Second, at intermediate fuel ratios
 ($\phi=0.6-0.8$) extended \textit{boomerang-shaped} patterns are
clearly visible. Note that these series lead to distributions very peaked near the mean, decline
rather rapidly, and have quite asymmetric left tails. In this interval our results also are in accordance with those of Daw \emph{et al.}~\cite{daw98,Green1999}. Third, at the
lowest equivalence ratio, $\phi=0.5$  a change of structure is observed, probably because of cycle misfires for this very poor mixture. Here the probability distribution is less peaked than a Gaussian, present a high standard deviation, and an asymmetric left tail.
Fourth, a closer inspection of Fig.~4 reveals some kind of
asymmetry around the diagonal, more pronounced as the fuel-air ratio
becomes lower. This is in accordance both with  experimental and model results by Daw \emph{et al.}~\cite{daw96,daw98,Green1999}. So, our stochastic scheme with the consideration of pseudo-random fluctuations on $l_t$ reproduces the characteristic return maps of this kind of systems from poor mixtures to over stoichiometric ones.

The results above refer to the
 the joint influence of the parameters
$u_t$ and $l_t$ on the CV-phenomena when both parameters are
considered linked by the empirical relations Eqs.~(\ref{ut}) and~(\ref{lt}). From now on our goal is double: on one
side, to analyze the influence of $u_t$ and $l_t$ but considered as
independent in the simulation model and, on the other side, to
analyze the influence of the third key parameter in the combustion
description, the location of ignition accounted for the
parameter $R_c$. To get this we have run the simulation for $2800$ consecutive cycles for the previously considered fuel-air equivalence ratios but taking these three parameters, one by one, as stochastic in nature.

First, we have considered only fluctuations on $l_t$,
taking the same distribution that at the beginning of this section. We  show the corresponding return maps in Fig.~5(a), that should be compared with Fig.~4. From this comparison we can conclude
that the main characteristics of the return maps obtained when $l_t$ and $u_t$ were considered as linked (noisy spots near stoichiometric conditions, boomerang-shaped structures at
intermediate fuel-air ratios and complex patterns at lower air-fuel
ratios) are already induced by the distribution of the
characteristic length $l_t$ when the other parameters are not stochastic. Differences between both figures only affect the dispersion of the boomerangs arms.

In regard to $u_t$, it is remarkable that it is not easy to find in the literature experimental data which allow to deduce stochastic distributions for this characteristic turbulent velocity with certainty. We have used the data in\cite{abdi2007} for the turbulence intensity and translated them to generate our log-normal distribution for the characteristic velocity, $u_t$, and checked slight changes in the distribution parameters within realistic intervals.
So, we assume a log-normal distribution
$LogN(\mu_{\log u_t},\sigma_{\log u_t})$ around the nominal value
$u_t^0=0.08\,\bar{u}_{i}(\rho_u/\rho_i)^{1/2}$ [Eq.~(\ref{ut})]  with standard deviation
$\sigma_{\log u_t}=0.02\,u_t^0$ and mean $\mu_{\log u_t}=\log (u_t^0)-(\sigma_{\log u_t}^2/2)$.
When    $u_t$ is the only stochastic parameter, the observed behaviors are not significantly altered (Fig.~5(b)):
boomerang-like arrangements are found only at low $\phi$,
although some sensitivity in the arms of the boomerangs is appreciable. Moreover, we have not found in the return maps new features respect to those generated with $l_t$.

Respect to the ignition
location, $R_c$, some authors~\cite{Stone1996} concluded from experiments  and models that displacement of the flame kernel during the early stages of combustion has a major part in the origination of cycle-by-cycle variations in combustion. So, it seems interesting to check this point from quasi-dimensional simulations.
It is remarkable that it is not straightforward to numerically evaluate the radius and area of the flame front when $R_c$ is displaced respect to the origin. We have generalized the studies by Bayraktar~\cite{bayraktar_prop} and Blizard~\cite{blizard_1974} to obtain numerical expressions for the radius and area of the flame front for whichever value of $R_c$ and at any time during flame evolution.
Second, we assume a Gaussian distribution with standard
deviation $\sigma_{R_c}=2.965\times 10^{-3}$ m, according to the results reported by
Beretta~\cite{beretta}. At sight of Fig.~5(c) it is clear that the influence of a stochastic component in $R_c$ is
limited to cause noisy spots at any fuel ratio, clear structures as boomerangs are not found at any fuel-air equivalence ratio. So, fluctuations on the displacement of the initial flame kernel seem not enough to reproduce characteristic patterns of CV.

Finally, we have checked what happens when the three parameters are simultaneously and independently introduced as stochastic in the simulations with the distributions mentioned above and no new features were discovered.
At sight of these results we can say that the observed heat release
behavior in terms of the fuel-air ratio is mainly due to stochastic
and non-additive variations of $l_t$ and $u_t$ and the non-linearity
induced by the dynamics of the combustion process, independently of
ignition location fluctuations.

Some physical consequences of cyclic variability are susceptible of a  clear interpretation in terms of the evolution of the fraction of burned gases in the chamber with time or with crank angle. This evolution has two main ingredients: the ignition delay (the interval until this fraction departs from zero)~\cite{letelier97} and the slope of that increase during combustion (see, for instance, Fig.~16 in~\cite{beretta}). These factors are mainly associated to eddies characteristic length $l_t$ and $u_t$, and the laminar speed $S_l$ (that in turn is a function of the fraction of residual gases, $y_r$, in the chamber, Eq.~(\ref{llama.laminar})). This last variable determines the memory effects from cycle to cycle, as noted by Daw \emph{et al}.~\cite{daw96,daw98}. Cyclic variability provokes cycle-to-cycle changes in the shape of the rate of burned gases, and this directly affects the evolution of pressure during the cycle, particularly its maximum value and its corresponding position. And consequently this leads to changes in the performance of the engine, \emph{i.e.}, in the power output and in the efficiency.

\section{Nonlinear analysis: Correlation integral and surrogates}

According with the conclusions of the preceding section from now on all the results we present were obtained considering fluctuations only on $l_t$, except when simulations were purely deterministic.
In order to get a better comprehension of the complexity of heat release fluctuations for different fuel ratio values, we next explore the nonlinear properties of the signals through the analysis of the correlation integral~\cite{grassberger1983,grassberger19832,kantz2004}. The correlation dimension analysis quantifies the self-similarity properties of a given sequence and is an important statistical tool which helps to evaluate the presence of determinism in the signal. We first reconstruct an auxiliary phase space by an embedding procedure. The correlation sum $C(\varepsilon,m)=2/[N(N-1)]\sum_{i=1}^N\sum_{j=i+1}^N\Theta(\varepsilon-\parallel\vec x_i-\vec x_j\parallel)$, where $\Theta$ is the Heaviside function, $\varepsilon$ is a distance and $\vec x_k$ are $m$-dimensional delay vectors, is computed for several values of $m$ and $\varepsilon$~\cite{hegger1999}. If $C(\varepsilon,m)\propto \varepsilon^{-d}$, the correlation dimension is defined as $d(\varepsilon)=d\log C(\varepsilon,m)/d\log\varepsilon$. To get a good estimation of $m$ we need to repeat the calculations for several values of $m$, the number of embedding dimensions. In this way we are able to select which one corresponds to the best option to characterize the dimensionality of the system. 
We notice that the application of correlation dimension analysis alone is not sufficient to differentiate between determinism and stochastic dynamics. To reinforce our calculations of correlation integral statistics, we consider a surrogate data method to verify that low values of correlation dimension of sequences are not a simple consequence of artifacts \cite{theiler1992}. To this end, two surrogate sets were considered. First, the original sequence was randomly shuffled to destroy temporal correlations. A second surrogate set was constructed by a phase randomization of the original sequence. The phase randomization was performed after applying the Fast Fourier Transform (FFT) algorithm to the original time series to obtain the amplitudes. Then the surrogate set was obtained by applying the inverse FFT procedure~\cite{theiler1992,schreiber2000}.

We apply the correlation dimension method to heat release sequences with $10^4$ cycles and for different values of the fuel ratio in the interval $0.5<\phi<0.7$. For larger fuel ratios we disregard the correlation dimension analysis since the signal shows small amplitude fluctuations around a stable value. According with the map-like characteristics observed in Fig.~4 we use a delay time equal to one. Figure~6 shows the correlation sum for two selected values of fuel ratio. We observe that for low values of $\phi$ ($\phi=0.5$, Figs.~6(a)) the correlation integral shows a power law behavior against $\varepsilon$ for several values of $m$. This behavior is best found by plotting the slope $d(\varepsilon)$ of $\log(\varepsilon,m)$ vs. $\log\varepsilon$. The inset shows this plot for values of $m$ in the range $1-5$. In this case there is a wide plateau of $\varepsilon$ which corresponds to the power law behavior. It is also clear that the height of the plateau increases with the embedding dimension with a reduction of amplitude of the power law region and some fluctuations for low values of $\varepsilon$.

In contrast, we observe in Fig.~6(b) that for intermediate fuel ratios ($\phi=0.65$) the scaling region only exists for values of $\varepsilon$ in the range $10<\varepsilon<10^2$. This is confirmed by the presence of a plateau for several values of $m$ (see the inset in Fig.~5(b)). Remarkably, the height of the plateau, $d(\varepsilon)$, almost does not change with the embedding dimension, indicating that the system can be characterized by a very low dimension. We also notice that for very small scales ($\varepsilon < 10$), a power law behavior can be also identified, but the scaling exponent increases to reach the value of the embedding dimension, suggesting to identify this region as a noisy regime.

Next, we use the surrogate data method to evaluate if the low value of correlation dimension observed for $\phi=0.65$ is not an artifact. We generated a randomly shuffled sequence and another set by a phase randomization of the original data. For both surrogate sets the correlation dimension was calculated in the same form and intervals as for original data. The results of these calculations are presented in Fig. 7.  We observe that for shuffled and phase randomized data the correlation dimension increases as the value of $m$ also increases, indicating that there is no saturation for correlation dimension (Fig. 7(a) and 7(b)). 

The findings about correlation dimension of original and surrogate data are summarized in the plots of Fig.~7(c). For low fuel ratio values, the correlation dimension does not saturate for high embedding values whereas for the intermediate fuel ratio  ($\phi=0.65$), correlation dimension saturates with the embedding dimension which is an indication of low dimensional dynamics. It is also interesting that when fluctuations on $l_t$ are not included in the simulations (deterministic case) the same situation is observed. The results for surrogate data (shuffled and phase randomized) reveal that no saturation of correlation dimension is observed, indicating that surrogate sequences differ from the original data and that the low dimensionality is probably related to the presence of determinism in heat release fluctuations.
In this sense, Scholl and Russ~\cite{scholl1999} have reported that deterministic patterns of CV are the consequence of incomplete combustion, which  occurs at $\phi=0.65$ in agreement with Heywood~\cite{heywood.intro}.

We present in Fig.~8 an extensive analysis of the evolution of the heat release with the fuel ratio between $\phi=0.5$ and $1.1$. The figure contains the results from the direct solution of the deterministic set of equations (black dots) and also from incorporating a stochastic component in $l_t$ (gray dots). For each value of $\phi$, the last $100$ simulated points of $250$ cycles runs are shown. We remark that under moderately lean fuel conditions ($\phi$ around $0.65$) our quasi-dimensional stochastic simulation does not show any period-2 bifurcation, as it happens in the theoretical model by Daw \emph{et al}.~\cite{daw96,daw98}. Our work predicts a very low dimensionality map in this region. Such conclusion makes sense since bifurcations have never been  reported in real car engines. 

On the other hand, the deterministic simulations do not lead to a fixed point as in the Daw's map, but to a variety of multiperiodic behavior for different fuel ratios. In particular, the inset of Fig.~8 shows the evolution of the normalized deterministic heat release, $Q_r/\bar Q_r$, ($\bar Q_r$ is the average heat release for each value of $\phi$) with fuel ratio between $0.5$ and $1.1$. We observe that for low and high fuel ratio values the number of fixed points is clearly smaller than those present for intermediate values of $\phi$, indicating a high multiperiodicity.  A deeper evaluation of the multiperiodicity observed in bifurcation-like plots for this system probably will deserve future studies.

\section{Discussion and conclusions} \label{s4}

We have developed a simulation scheme in order to reproduce the experimentally observed fluctuations of heat release in an Otto engine. The model relies on the first principle of thermodynamics applied to open systems that allows to build up a set of first order differential equations for pressure and temperature inside the cylinder. Our model includes a detailed chemistry of combustion that incorporates the presence of exhaust gases in the fresh mixture of the following cycle. The laminar combustion speed depends on the mole fraction of residual gases in the chamber through a phenomenological law~\cite{bayraktar_LPG}. Combustion requires a particular model giving the evolution of the masses of unburned and burned gases in the chamber as a function of time. We consider a model where inside the approximately spherical flame front there are unburned eddies of typical length $l_t$. The model incorporates two other parameters, the position of the kernel of combustion, $R_c$, and a convective characteristic velocity $u_t$ (related with turbulence), that could be important in order to reproduce the observed cycle-to-cycle variations of heat release. Actually, we have checked the influence of each of those parameters when they are considered as stochastic. It is possible from experimental results in the literature to build up log-normal probability distributions for $l_t$ and $u_t$, and Gaussian for $R_c$. The results of our simulations with these stochastic distributions show that the consideration of $l_t$ or $u_t$ as stochastic is essential to reproduce experiments. These parameters can be considered as independent or related through an empirical correlation. Both possibilities lead to similar results. On the contrary a stochastic component on $R_c$ only provokes noisy spots in the return maps for any fuel ratio.

Moreover, an interesting evolution of the temporal series of heat release is obtained when the fuel ratio, $\phi$, is considered as a parameter. The dependence on $\phi$ of the main statistical parameters of heat release sequences for real engines are reproduced~\cite{daw98,sen2010}. Specially interesting is the evolution of skewness and kurtosis.
Return maps also have a rich behavior: from noisy unstructured clusters (\emph{shotgun} patterns) at high fuel ratios to boomerang asymmetric motives at intermediate $\phi$. This behavior was obtained before from mathematical models and also from experiments~\cite{daw96,daw98}.

The correlation integral analysis reveals that, for a certain range of the distance $\varepsilon$, the dimensionality of the system is small only for intermediate fuel ratios. We have additionally used the surrogate data method to compare the findings of dimensionality between original simulated and randomized data.  Our results support the fact that the low dimensionality observed for intermediate values of the fuel-air equivalence ratio (around $\phi=0.65$), is related to the presence of determinism in heat release fluctuations. When a systematic study of the evolution of the heat release calculated from the stochastic simulations with $\phi$ is performed, bifurcations were not found. This is in agreement with real engines, where to our knowledge period doubling bifurcations have not been found.

In summary, our numerical model incorporates the main physical and chemical ingredients necessary to reproduce the
evolution with fuel ratio of heat release CV. We have particularly analyzed the influence on CV phenomena of three basic
parameters governing the turbulent combustion process: length of the
unburned eddies, the turbulent intensity, and the location of the
ignition site. The influence of the two first ones seems to be
basic in the observed structures of the heat release time
series.
The control of these intermittent fluctuations of heat release could result
 in heat engines with greater mean values of
power and efficiency. Particular rich behavior of heat release at certain low and intermediate fuel ratios deserves a detailed study of its non-linear dynamics. Work along these lines is in progress.

\section*{Acknowledgements}

Authors acknowledge financial support  from \emph{Junta de Castilla y León} under Grant SA054A08.
P.L.C.-R. acknowledges a pre-doctoral grant from Grupo
Santander-Universidad de Salamanca. L.G.-V. and F.A.-B. thank COFAA-IPN, EDI-IPN and Conacyt (49128-26020), México.

\clearpage
\bibliographystyle{elsarticle-num}

\clearpage

\begin{figure}[ht!]
\centering
  \includegraphics[width=0.8\textwidth]{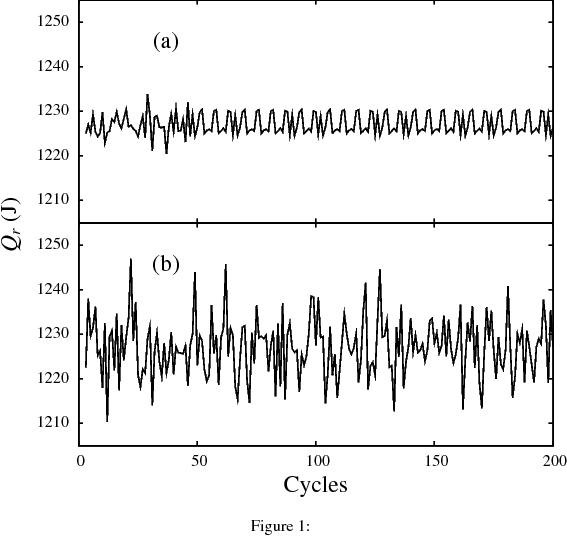}
  \caption{(a) Heat release time evolution as obtained from the simulation without the consideration of any stochastic component for $\phi=1.0$. (b) Heat release time series, $Q_{r}$, obtained with a log-normal distribution (see text) for the characteristic length for unburned eddies during combustion, $l_t$ .}
\end{figure}

\begin{figure}[ht!]
\centering
  \includegraphics[width=0.8\textwidth]{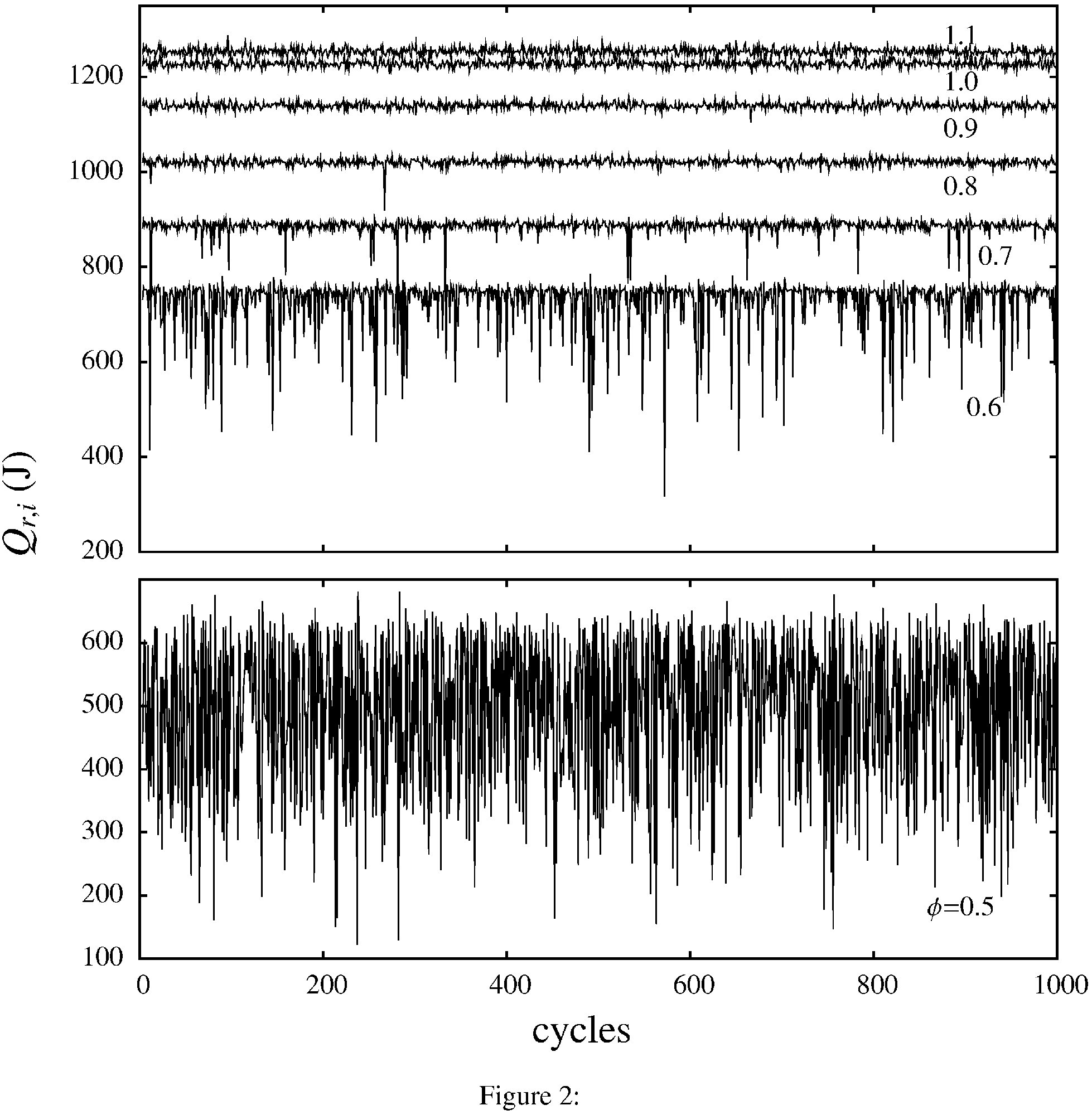}
  \caption{Heat release time series, $Q_{r}$, obtained with a log-normal distribution (see text) for the characteristic length of unburned eddies during combustion, $l_t$. Results are shown for several fuel-air equivalence ratio ($\phi$ increases from bottom to top from $\phi=0.5$ to $1.1$).}
\end{figure}

\begin{figure}[ht!]
\centering
  \includegraphics[width=0.8\textwidth]{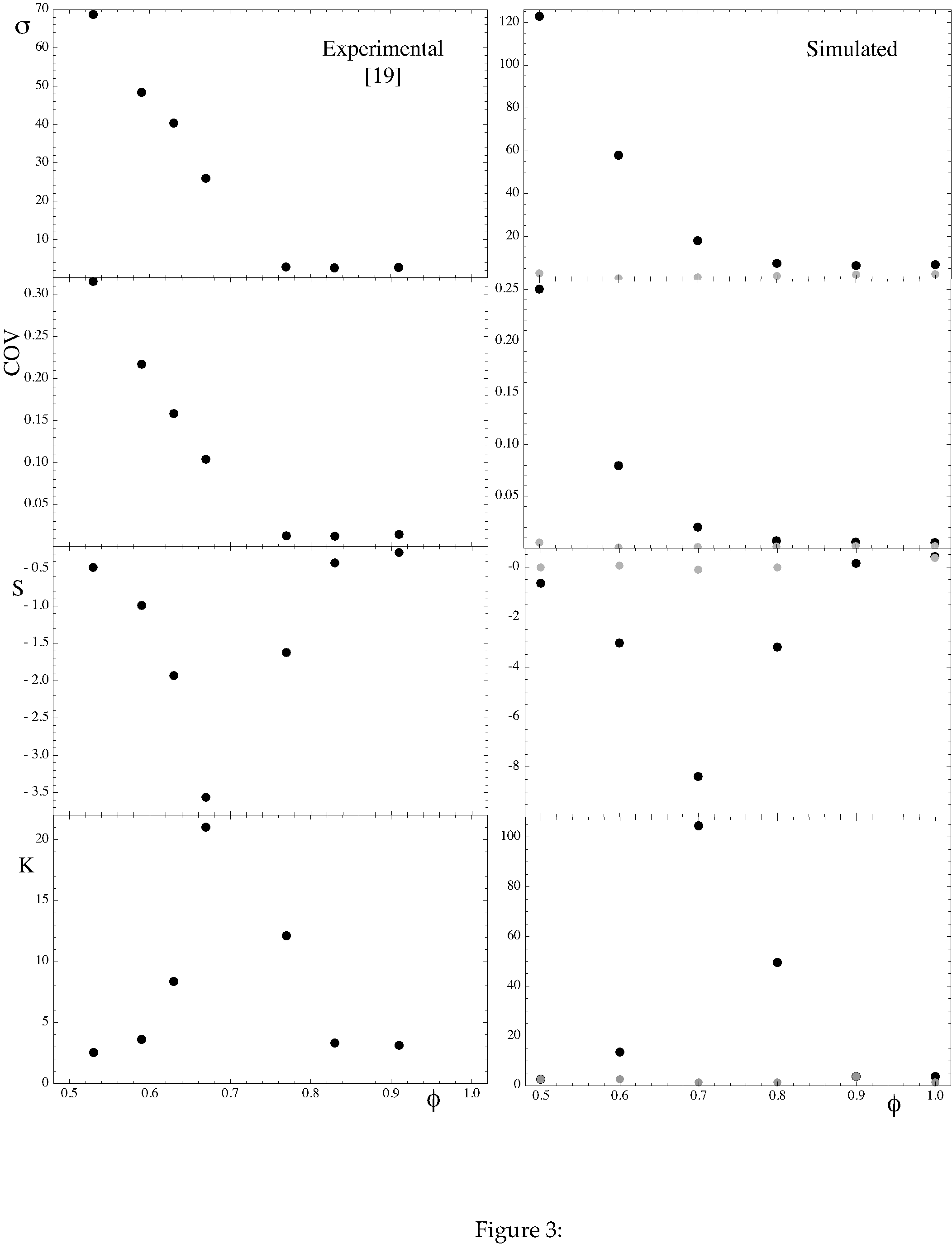}
  \caption{Evolution of some statistical parameters with the fuel ratio. Left panel, experiments~\cite{daw98,sen2010} and right panel,  stochastic simulation results.}
\end{figure}

\begin{figure}[ht!]
\centering
  \includegraphics[width=0.8\textwidth]{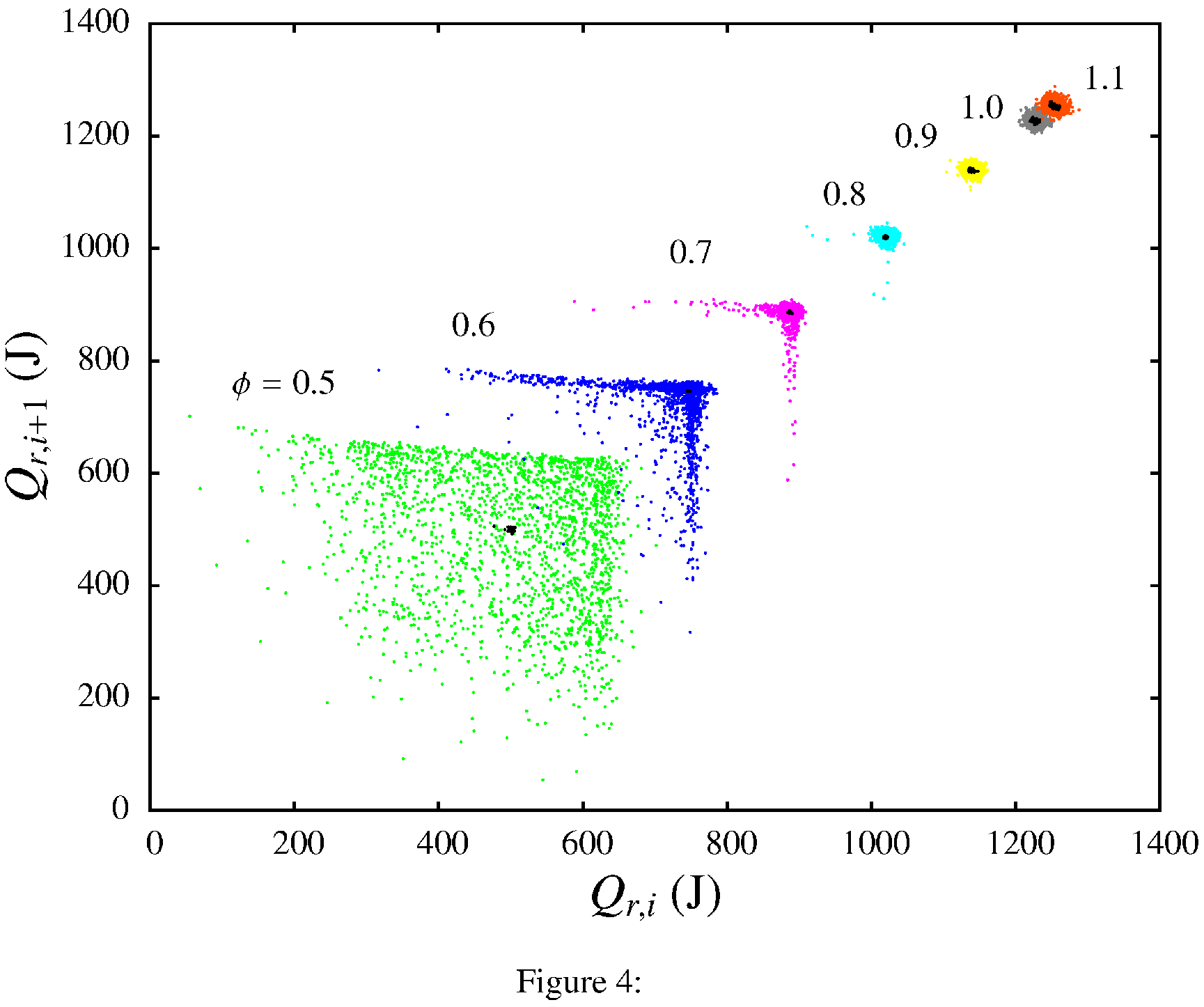}
  \caption{Heat release return maps after 2800 simulation runs for the same  values of $\phi$ that in Fig.~2 when $l_t$ is considered as stochastic. For each $\phi$ the central dark kernel corresponds to the deterministic simulation.}
\end{figure}

\begin{figure}[ht!]
\centering
  \includegraphics[width=0.8\textwidth]{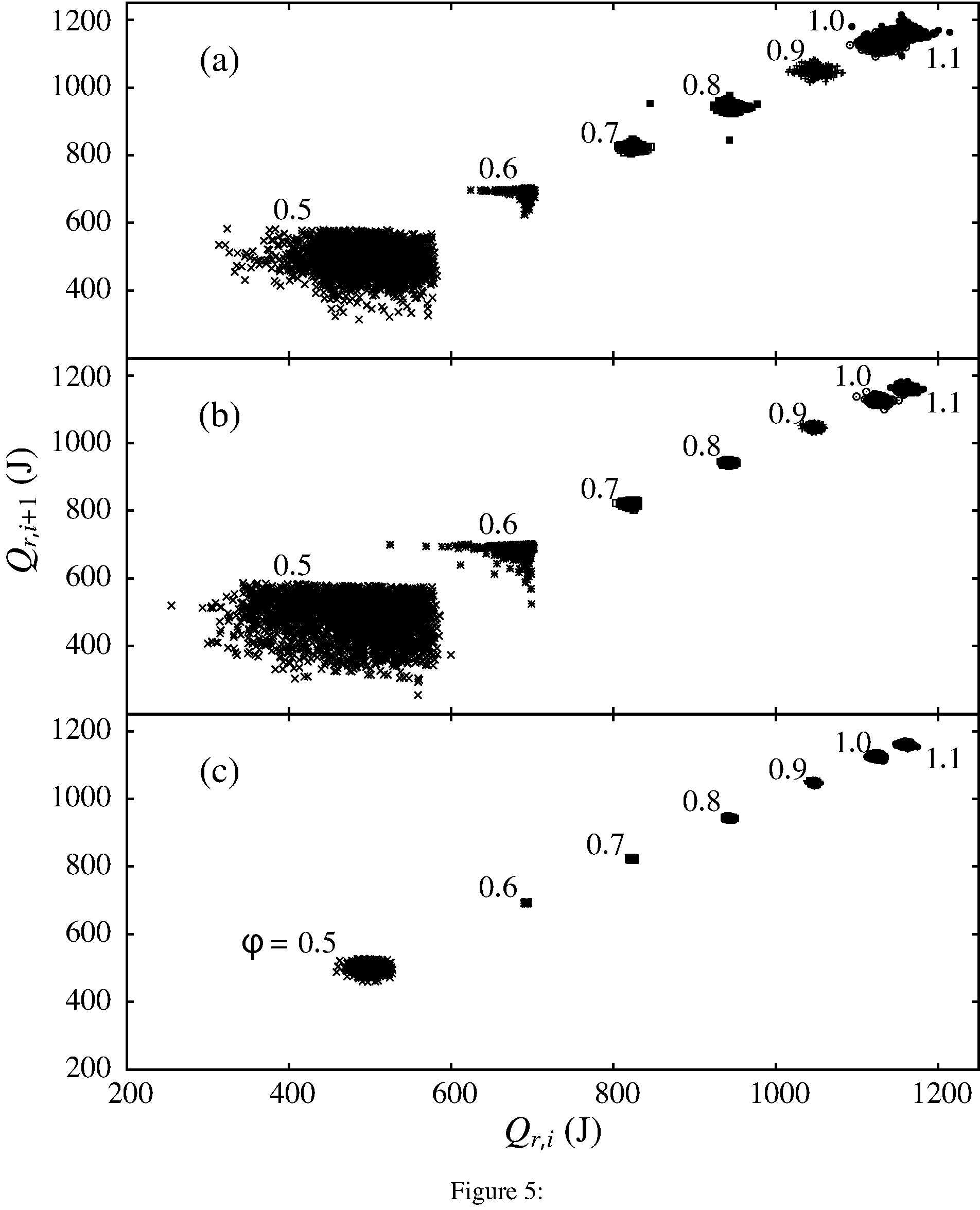}
  \caption{Heat release return maps obtained when the three basic parameters of combustion, $l_t$, $u_t$, and $R_c$ are considered as independent and one-by-one stochastic. (a) Fluctuations are only introduced in $l_t$ (see text for details on the probability distribution); (b) only $u_t$ fluctuates, and (c) only $R_c$ is considered as stochastic.}
\end{figure}

\begin{figure}[ht!]
\centering
  \includegraphics[width=0.5\textwidth]{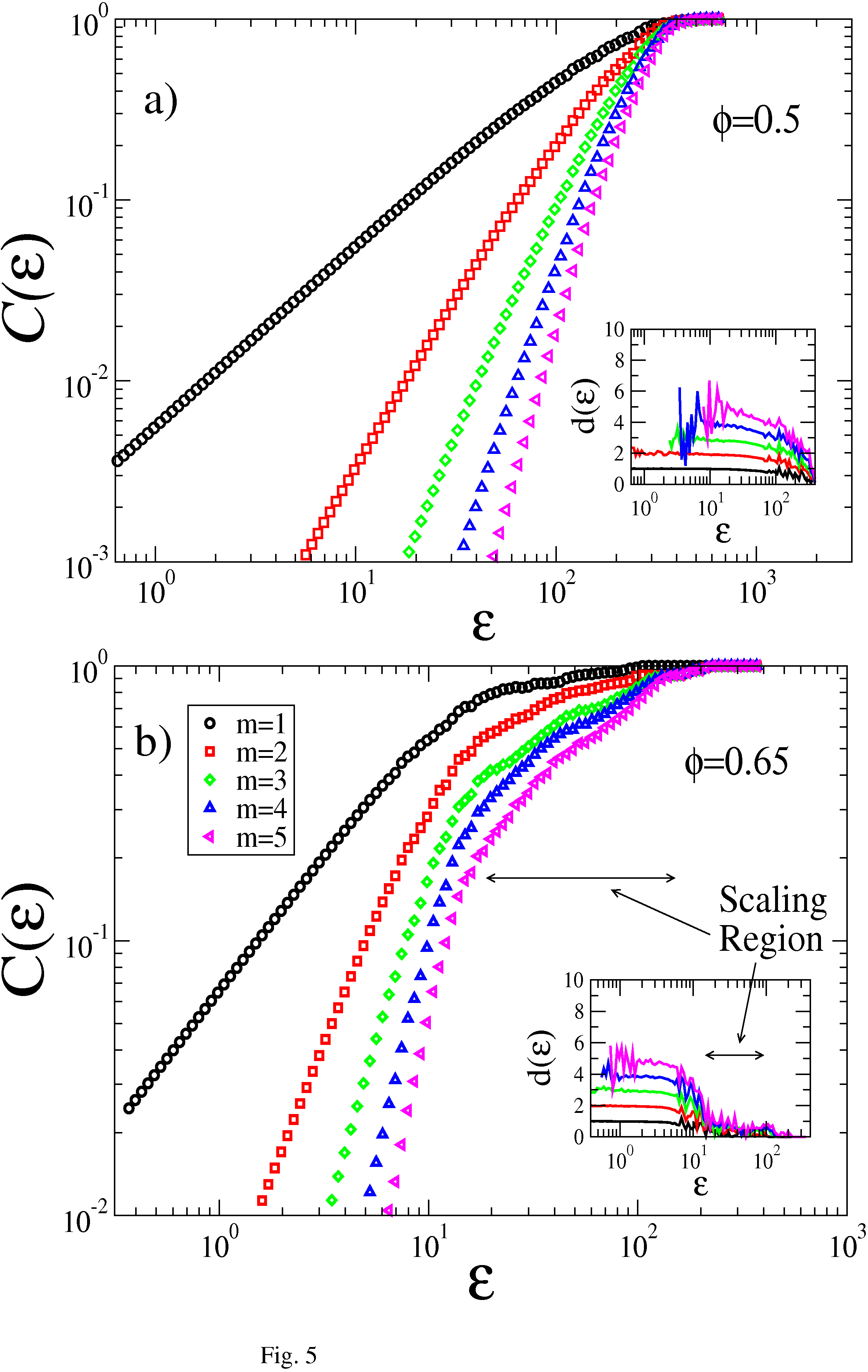}
  \caption{Correlation integral $C(\varepsilon)$ as a function of the distance $\varepsilon$ for several embedding dimensions $m$ and two values of $\phi=0.5$ and $\phi=0.65$. The insets show in each case logarithmic plots of the correlation dimension vs. $\varepsilon$.}
\end{figure}

\begin{figure}[ht!]
\centering
  \includegraphics[width=0.5\textwidth]{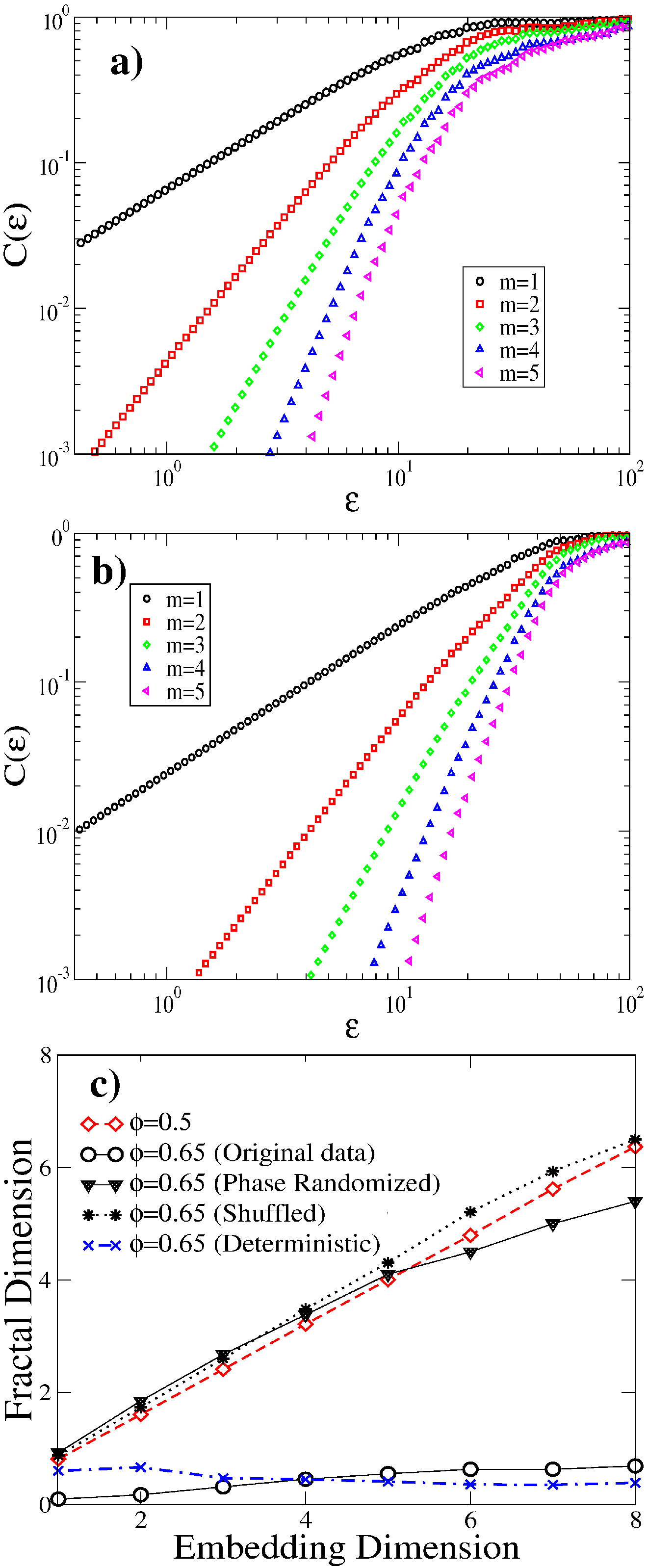}
  \caption{Correlation integral for surrogate data sets for $\phi=0.65$ obtained from (a) shuffling  and (b) phase randomization procedure. (c) Correlation dimension in terms of the embedding dimension for low and
intermediate values of the fuel ratio. We show the cases of original data ($\phi=0.5$ and $0.65$) and the results for shuffled  and phase randomized data together with the deterministic case.}
\end{figure}

\begin{figure}[ht!]
\centering
  \includegraphics[width=0.8\textwidth]{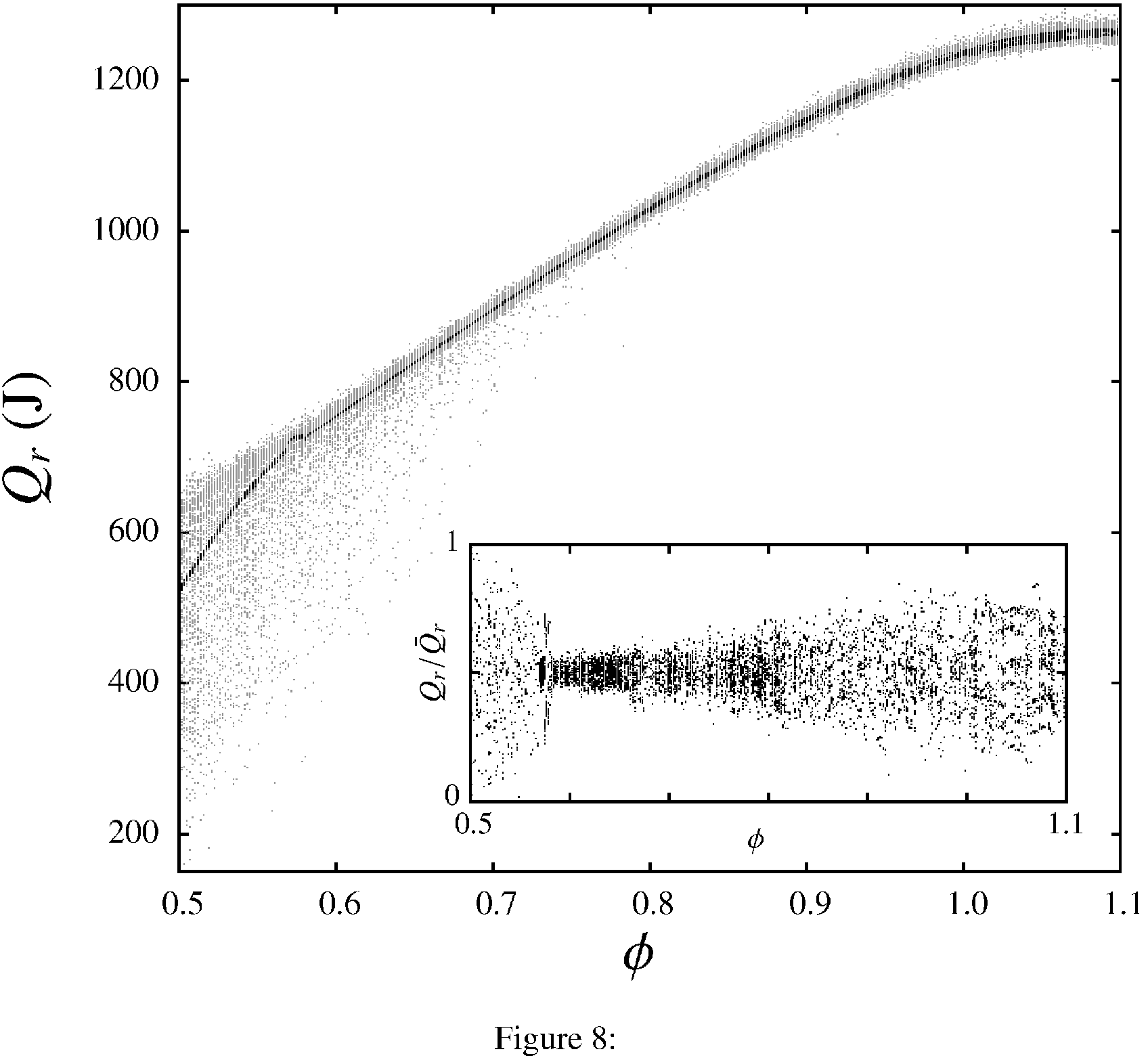}
  \caption{Evolution of heat release  time series as a function of $\phi$. Dark dots correspond to the deterministic simulation, and the inhomogeneous gray dots distribution to the stochastic approach. The inset shows the normalized heat release, $Q_r/\bar Q_r$ obtained from the deterministic simulations ($\bar Q_r$ is the average heat release for each fuel ratio).}
\end{figure}

\clearpage

\begin{table}[!ht]
\centering
\caption{Statistical parameters of the heat release temporal series,
considering the deterministic model, for several fuel ratio values: mean
value, $\mu$,
standard deviation, $\sigma$, coefficient of covariance, $COV$,
skewness, $S$, and kurtosis, $K$.}
\label{statdet}
\medskip
\small
\begin{tabular}{cccccccc}
\hline\hline
$\phi$  & $0.5$ & $0.6$ & $0.7$ & $0.8$ & $0.9$ & $1.0$ & $1.1$
\\
\hline
$\mu$ (J)  & $500.84$ & $746.19$ & $886.82$ & $1019.73$ &
$1138.79$ & $1227.20$ & $1253.32$\\
$\sigma$ (J) & $2.593$ & $0.320$ & $0.785$ & $1.463$ & $1.969$
& $2.249$ & $2.255$\\
$COV$ ($\times10^{-3}$)  &  $5.177$ &  $0.428$ & $0.885$ &
$1.435$
& $1.729$ & $1.832$ & $1.799$\\
$S$  & $-0.004$ & $0.057$ & $-0.106$ & $-0.013$ & $1.317$ &
$0.376$ & $-0.062$\\
$K$  & $2.485$ & $2.638$ & $1.332$ & $1.326$ & $3.610$ &
$1.361$ & $3.458$\\
\hline\hline
\end{tabular}
\end{table}

\begin{table}[!ht]
\centering
\caption{Statistical parameters of the heat release temporal series
represented in Fig.~2 and obtained considering stochastic fluctuations in $l_t$, for several fuel ratio values.}
\label{stat}
\medskip
\small
\begin{tabular}{cccccccc}
\hline\hline
$\phi$  & $0.5$ & $0.6$ & $0.7$ & $0.8$ & $0.9$ & $1.0$ & $1.1$\\
\hline
$\mu$ (J)  & $492.50$ & $726.23$ & $885.43$ & $1020.24$ &
$1139.51$
& $1227.16$ & $1253.71$\\
$\sigma$ (J)  & $122.966$ & $57.872$ & $17.937$ & $7.333$ &
$6.326$ &
$6.663$ & $7.093$\\
$COV$ ($\times10^{-3}$)  & $250.0$ & $79.7$ & $20.3$ & $7.2$ &
$5.6$ & $5.4$ &
$5.7$ \\
$S$  & $-0.643$ & $-3.028$ & $-8.388$ & $-3.194$ & $0.147$ &
$0.423$ & $0.335$\\
$K$  & $2.495$ & $13.455$ & $104.475$ & $49.491$ & $3.579$ &
$3.710$ & $3.454$\\
\hline\hline
\end{tabular}
\end{table}

\end{document}